\documentstyle[12pt,axodraw]{article}
\def\lesssim{\mathrel{\mathpalette\vereq<}}
\def\gtrsim{\mathrel{\mathpalette\vereq>}}

\def\vev#1{\left\langle#1\right\rangle}
\def\lsim{\lesssim}
\def\gsim{\gtrsim}

\makeatletter
\def\vereq#1#2{\lower3pt\vbox{\baselineskip1.5pt \lineskip1.5pt
\ialign{$\m@th#1\hfill##\hfil$\crcr#2\crcr\sim\crcr}}}
\makeatother

\begin{document}

\begin{titlepage}

\begin{flushright}
SNS-PH-01-13 \\
UCB-PTH-01/22 \\
LBNL-48256 \\
\end{flushright}

\vskip 1.5cm

\begin{center}
{\Large \bf  Softly Broken Supersymmetric Desert \\ 
             from Orbifold Compactification}

\vskip 1.0cm

{\bf
Riccardo Barbieri$^{a}$,
Lawrence J.~Hall$^{b,c}$,
Yasunori Nomura$^{b,c}$
}

\vskip 0.5cm

$^a$ {\it Scuola Normale Superiore and INFN, Piazza dei Cavalieri 7, 
                 I-56126 Pisa, Italy}\\
$^b$ {\it Department of Physics, University of California,
                 Berkeley, CA 94720, USA}\\
$^c$ {\it Theoretical Physics Group, Lawrence Berkeley National Laboratory,
                 Berkeley, CA 94720, USA}

\vskip 1.0cm

\abstract{A new viewpoint for the gauge hierarchy problem is proposed: 
compactification at a large scale, $1/R$, leads to a low energy
effective theory with supersymmetry softly broken at a much lower
scale, $\alpha / R$. The hierarchy is induced by an extremely small
angle $\alpha$ which appears in the orbifold compactification boundary 
conditions. The same orbifold boundary conditions break Peccei-Quinn
symmetry, leading to a new solution to the $\mu$ problem.
Explicit 5d theories are constructed with gauge groups 
$SU(3) \times SU(2) \times U(1)$ and $SU(5)$, with matter in the bulk
or on the brane, which lead to the (next-to) minimal supersymmetric 
standard model below the compactification scale. In all cases the soft
supersymmetry-breaking and $\mu$ parameters originate from bulk 
kinetic energy terms, and are highly constrained. The supersymmetric 
flavor and CP problems are solved.}

\end{center}
\end{titlepage}

\section{Introduction}
\label{sec:intro}

Data from precision electroweak experiments, which includes evidence
in favor of a light Higgs boson, have made weak scale supersymmetry
the leading candidate for a theory beyond the standard model. Weak
scale supersymmetry provides a solution to the gauge hierarchy
problem, a radiative electroweak symmetry breaking mechanism with a
light Higgs boson, and a successful prediction for the weak mixing
angle.  The critical question for weak scale supersymmetry is: what 
breaks supersymmetry? In many schemes this is accomplished in 4d by 
the dynamics of some new strong gauge force. In this paper we explore 
an alternative possibility: the breaking of supersymmetry by boundary 
conditions in compact extra dimensions \cite{Scherk:1979ta}. 
While such a mechanism has been known for many years, 
it has rarely been applied to realistic models.  Models which have 
been constructed \cite{Antoniadis:1990ew, Antoniadis:1993fh, 
Pomarol:1998sd, Antoniadis:1999sd, Delgado:1999qr, Barbieri:2001vh, 
Arkani-Hamed:2001mi, Delgado:2001si}, have taken the view that 
the compactification scale $1/R$ is of order a TeV, and that beneath 
this scale supersymmetry is broken. Thus the picture is of a transition 
at $1/R$ from a $d>4$ supersymmetric theory directly to a $d=4$
non-supersymmetric effective theory. There is never an energy interval
in which there is an effective 4d supersymmetric field theory. Such 
schemes are extremely exciting, as they predict that both Kaluza-Klein 
(KK) modes and superpartners will be discovered by colliders at the TeV
scale. However, in these schemes supersymmetry is apparently not
related to the gauge hierarchy problem, and logarithmic gauge coupling
unification is not possible.

In this paper we demonstrate that there is an alternative
implementation of boundary condition supersymmetry breaking:
the boundary conditions may involve very small dimensionless
parameters, $\alpha$, so that supersymmetry is broken at $\alpha/R$
rather than $1/R$. In this scheme the transition at scale $1/R$ is
from a 5d supersymmetric theory to a 4d supersymmetric theory with
highly suppressed supersymmetry breaking interactions.  In the
energy interval from $1/R$ to $\alpha/R$ physics is described by 
a softly broken 4d supersymmetric theory, such as the minimal
supersymmetric standard model. This new viewpoint gives a new origin
for the soft supersymmetry-breaking parameters in terms of orbifold
compactification boundary conditions at very high energies.
For $1/R$  sufficiently high, supersymmetry is relevant for solving
the gauge hierarchy problem and logarithmic gauge coupling unification 
may occur. We do not claim this as a new solution to the gauge hierarchy
problem, as we have not understood why the parameter $\alpha$ is so
small, but we are hopeful that this new view of the problem may lead
to a new solution. 

In this paper we restrict our analysis to the simplest case of a
single compact extra dimension, in which case there is a unique
parameter, $\alpha$, in the orbifold boundary condition which breaks
supersymmetry \cite{Barbieri:2001dm}. This parameter arises as 
a twisting of the fields under a translation symmetry of the extra 
coordinate. In higher dimensions there will be further parameters. 
In 5d, with a single extra dimension, even with arbitrary
gauge and matter content of the theory, we are guaranteed that the 
soft supersymmetry-breaking parameters depend on only two parameters: 
$\alpha$ and $1/R$.

Realistic theories with supersymmetry in 4d must have two Higgs
chiral multiplets, vectorlike with respect to the standard model gauge
group. Any such theory must address why these Higgs doublets have
survived to the low energy effective theory --- why did they not get a
large gauge invariant mass? An obvious answer is that they are
protected by a global symmetry $G_H$ --- either Peccei-Quinn 
symmetry or $R$ symmetry. However, in this case one must address 
why the level of $G_H$ breaking scale is comparable to that of 
supersymmetry breaking, as required by phenomenology. We will solve 
these problems as follows. The Higgs fields propagate in the bulk and 
are forbidden to have a bulk mass term by orbifold symmetry and $G_H$. 
This same orbifold symmetry is such that, on making a KK expansion, 
there are two zero-mode Higgs doublets. The breaking of $G_H$ is 
accomplished by an orbifold boundary condition involving a 
dimensionless twisting parameter $\gamma$. The relevant orbifold 
symmetry is precisely the same translation of the extra coordinate 
that breaks supersymmetry and hence one naturally expects 
$\gamma \approx \alpha$. There is a unification of the origin of 
supersymmetry and $G_H$ breaking, providing a novel solution to the 
$\mu$ problem. 

In the next section we study the case that the gauge group is 
$SU(3) \times SU(2) \times U(1)$, so that the orbifold breaks
supersymmetry and global symmetry, but not gauge symmetry. We obtain
the form of the supersymmetry and $G_H$ breaking interactions, and 
study radiative electroweak symmetry breaking, both for the case of 
quarks and leptons on a brane and in the bulk. 

In section \ref{sec:su5} we study the case that the gauge group 
is $SU(5)$, and that the $SU(5)$ gauge symmetry is broken to that of 
the standard model by the same orbifold translation symmetry that 
breaks both supersymmetry and $G_H$.  The gauge symmetry breaking is 
induced by a set of parities and does not involve any small 
parameter, and hence occurs at the scale of $1/R$. In the limit 
$\alpha, \gamma \rightarrow 0$ this theory is the same as that 
studied in Refs.~\cite{Kawamura:2000ev, Hall:2001pg}. Including
non-zero values for $\alpha, \gamma$ allows a unified view of gauge, 
global and supersymmetry breaking.

\section{SU(3)$\times$SU(2)$\times$U(1) Model}
\label{sec:mssm}

In this section, we construct 5d theories which, at low energies, reduce 
to the minimal supersymmetric standard model with specific forms of the 
soft supersymmetry-breaking parameters.

\subsection{The model}
\label{subsec:mssm-brane}

The gauge group is taken to be $SU(3) \times SU(2) \times U(1)$.  
The 5d gauge multiplet ${\cal V} = (A^M, \lambda, \lambda', \sigma)$ 
is decomposed into a vector superfield $V=(A^\mu, \lambda)$ and 
a chiral superfield $\Sigma=(\sigma+iA^5, \lambda')$ under 4d $N=1$ 
supersymmetry.  We also introduce two Higgs hypermultiplets 
${\cal H}_i = (h_i, h^{c\dagger}_i, \tilde{h}_i, \tilde{h}^{c\dagger}_i)$
$(i=1,2)$ in the 5d bulk.   Under 4d $N=1$ supersymmetry, each of them 
is decomposed into two chiral superfields $H_i = (h_i, \tilde{h}_i)$ and 
$H^c_i = (h^c_i, \tilde{h}^c_i)$, where $H_i$ and $H_i^c$ have conjugated 
transformations under the gauge group.  A large bulk mass term is 
forbidden by imposing a global symmetry $G_H$ under which $\tilde{h}_1$ 
and $\tilde{h}^c_2$ transform in the same way.

The fifth dimension is compactified on the $S^1/Z_2$ orbifold, which is 
constructed by two identifications $y \leftrightarrow -y$ and 
$y \leftrightarrow y+2\pi R$.  Under the first identification, 
$y \leftrightarrow -y$, the gauge and Higgs fields are assumed to obey 
the following boundary conditions:
\begin{eqnarray}
  \pmatrix{V \cr \Sigma}(x^\mu,-y) &=& \pmatrix{V \cr -\Sigma}(x^\mu,y), \\
  \pmatrix{H_1 & H_2 \cr H_1^{c\dagger} & H_2^{c\dagger}}(x^\mu,-y) 
  &=& \pmatrix{H_1 & -H_2 \cr -H_1^{c\dagger} & H_2^{c\dagger}}(x^\mu,y).
\end{eqnarray}
This leaves only 4d $N=1$ $SU(3) \times SU(2) \times U(1)$ vector 
superfields and two Higgs chiral superfields $H_1$ and $H_2^c$ as 
zero-modes, upon compactifying to $S^1/Z_2$.  All the other states have 
masses of order $1/R$.  We also impose the following boundary conditions 
under $y \leftrightarrow y+2\pi R$:
\begin{eqnarray}
  A^M(x^\mu,y+2\pi R) &=& A^M(x^\mu,y), 
\label{eq:bc-T-A}\\
  \pmatrix{\lambda \cr \lambda'}(x^\mu,y+2\pi R) &=& 
	e^{-2\pi i \alpha \sigma_2} 
	\pmatrix{\lambda \cr \lambda'}(x^\mu,y), 
\label{eq:bc-T-lambda}\\
  \sigma(x^\mu,y+2\pi R) &=& \sigma(x^\mu,y), 
\label{eq:bc-T-sigma}\\ \nonumber\\
  \pmatrix{h_1 & h_2 \cr h_1^{c\dagger} & h_2^{c\dagger}}(x^\mu,y+2\pi R) 
	&=& e^{-2\pi i \alpha \sigma_2} 
	\pmatrix{h_1 & h_2 \cr h_1^{c\dagger} & h_2^{c\dagger}}
	e^{2\pi i \gamma \sigma_2} (x^\mu,y), 
\label{eq:bc-T-h}\\
  \pmatrix{\tilde{h}_1 & \tilde{h}_2 \cr 
	\tilde{h}_1^{c\dagger} & \tilde{h}_2^{c\dagger}}(x^\mu,y+2\pi R) 
	&=& \pmatrix{\tilde{h}_1 & \tilde{h}_2 \cr 
	\tilde{h}_1^{c\dagger} & \tilde{h}_2^{c\dagger}}
	e^{2\pi i \gamma \sigma_2} (x^\mu,y).
\label{eq:bc-T-tildeh}
\end{eqnarray}
where $\alpha$ and $\gamma$ are continuous parameters, and 
$\sigma_{1,2,3}$ are the Pauli spin matrices.  Note that this is the most 
general boundary condition under the $S^1/Z_2$ compactification with the 
present matter content \cite{Barbieri:2001dm}; $\alpha$ and $\gamma$ 
parameterize $U(1)$ rotations which are subgroups of the $SU(2)_R$ 
symmetry and flavor $SU(2)_H$ symmetry of the 5d action, respectively.
The boundary conditions Eqs.~(\ref{eq:bc-T-A} -- \ref{eq:bc-T-tildeh})
provide $G_H$ breaking, and induce the soft supersymmetry-breaking 
masses of $O(\alpha/R)$ and the supersymmetric mass for the Higgs fields 
of $O(\gamma/R)$ as we will see below. 

This theory was first introduced with the viewpoint that $\alpha$
and $\gamma$ are of order unity, so that the theory below the
compactification scale is the standard model rather than the
supersymmetric standard model \cite{Pomarol:1998sd}. This led to an 
emphasis of the phenomenology of the case $\gamma = \alpha$, 
since only in this limit was a light Higgs doublet obtained 
\cite{Delgado:1999qr}. Here we stress that we are interested 
in the very different viewpoint of $\alpha$ and $\gamma$ being 
extremely small, so that there is a large energy interval in which 
the theory is the minimal supersymmetric standard model.

We now consider the mode expansions for the various fields under the 
boundary conditions Eqs.~(\ref{eq:bc-T-A} -- \ref{eq:bc-T-tildeh}).
Non-trivial decompositions are required for the gauginos, Higgs bosons 
and Higgsinos.  They are given by
\begin{eqnarray}
  \pmatrix{\lambda \cr \lambda'}(x^\mu,y) &=& 
	\sum_{n=0}^{\infty} e^{-i \alpha \sigma_2 y / R} 
	\pmatrix{\lambda_n \cos[ny/R] \cr 
	\lambda'_n \sin[ny/R]},
\label{eq:KK-lambda}\\
  \pmatrix{h_1 & h_2 \cr h_1^{c\dagger} & h_2^{c\dagger}}(x^\mu,y) 
	&=& \sum_{n=0}^{\infty} e^{-i \alpha \sigma_2 y / R} 
	\pmatrix{h_{1n} \cos[ny/R] & h_{2n} \sin[ny/R] \cr 
	h_{1n}^{c\dagger} \sin[ny/R] & h_{2n}^{c\dagger} \cos[ny/R]}
	e^{i \gamma \sigma_2 y / R}, 
\label{eq:KK-h}\\
  \pmatrix{\tilde{h}_1 & \tilde{h}_2 \cr 
	\tilde{h}_1^{c\dagger} & \tilde{h}_2^{c\dagger}}(x^\mu,y) 
	&=& \sum_{n=0}^{\infty} \pmatrix{\tilde{h}_{1n} \cos[ny/R] & 
	\tilde{h}_{2n} \sin[ny/R] \cr 
	\tilde{h}_{1n}^{c\dagger} \sin[ny/R] & 
	\tilde{h}_{2n}^{c\dagger} \cos[ny/R]}
	e^{i \gamma \sigma_2 y / R},
\label{eq:KK-tildeh}
\end{eqnarray}
where $\lambda_n, \lambda'_n, h_{in}, h_{in}^c, \tilde{h}_{in}$ 
and $\tilde{h}_{in}^c$ are 4d fields.
On substituting these mode expansions into the 5d action and integrating 
out the heavy modes with masses of $O(1/R)$, we obtain the 4d effective 
Lagrangian below the scale of $1/R$.  It contains only the 
$SU(3) \times SU(2) \times U(1)$ vector superfields and two Higgs chiral 
superfields $H_1$ and $H_2^c$, which we define as $H_u \equiv H_{1,n=0}$ 
and $H_d \equiv H_{2,n=0}^c$.  In addition to the kinetic terms for these 
fields, there are mass terms coming from the boundary conditions 
Eqs.~(\ref{eq:bc-T-A} -- \ref{eq:bc-T-tildeh}),
\begin{eqnarray}
  {\cal L} &=& - \frac{1}{2} \frac{\alpha}{R} 
	(\lambda^a_0 \lambda^a_0 + {\rm h.c.}) \nonumber\\
  && - \left( \frac{\alpha^2}{R^2} + \frac{\gamma^2}{R^2} \right)
	(|h_u|^2 + |h_d|^2) 
	+ 2 \frac{\alpha\gamma}{R^2} (h_u h_d + {\rm h.c.}) \nonumber\\
  && - \frac{\gamma}{R} (\tilde{h}_u \tilde{h}_d + {\rm h.c.}),
\label{eq:higgs-pot}
\end{eqnarray}
where various fields are canonically normalized in 4d, and $a$ 
runs over $SU(3)$, $SU(2)$ and $U(1)$.  We find that 
the gaugino masses, soft supersymmetry-breaking masses for 
the Higgs bosons, the supersymmetric Higgs mass ($\mu$ term) and 
the holomorphic supersymmetry-breaking Higgs mass ($\mu B$ term) are 
generated.  Therefore, the low energy theory has the structure of the 
minimal supersymmetric standard model with various relations on the soft 
supersymmetry-breaking parameters.  In particular, it determines the 
sign of the $\mu$ parameter in the basis where $\vev{h_u}, 
\vev{h_d} > 0$.  The interactions in Eq.~(\ref{eq:higgs-pot}) are 
such that, using conventional notation, the sign of $\mu$ is negative.
(In the conventional notation, a negative $\mu$ leads to a stronger 
constraint from $b \rightarrow s \gamma$.)  An interesting point is that 
the sizes of $\alpha$ and $\gamma$ are expected to be the same order, 
since the $U(1)$ symmetry used to twist the boundary condition under 
$y \leftrightarrow y+2\pi R$ is a generic linear combination of 
two $U(1)$ symmetries, $U(1)_R \subset SU(2)_R$ and $U(1)_H \subset 
SU(2)_H$, associated with $\alpha$ and $\gamma$. Therefore, this 
theory provides a natural solution to the $\mu$ problem.

Orbifold breaking has led to a soft rather than hard breaking of 
supersymmetry.  When the KK mode expansions of Eqs.~(\ref{eq:KK-lambda}, 
\ref{eq:KK-h}) are substituted into the kinetic energy, the $y$
derivatives give $\alpha/R$ and result in soft operators, while the 4d
derivatives do not lead to supersymmetry breaking.\footnote{
We thank Alex Pomarol for pointing out our previous error on this point.} 
The supersymmetry-breaking parameter $\alpha/R$ drops out of the 
4d kinetic terms (kinetic terms with $\partial_\mu$) because of 
$SU(2)_R$ invariance. It does not drop out of the bulk kinetic term 
(kinetic terms with $\partial_y$) because $SU(2)_R$ is a global symmetry, 
and the phase factors in Eqs.~(\ref{eq:KK-lambda}, \ref{eq:KK-h}) are 
$y$ dependent. The resulting supersymmetry breaking interactions are soft
by dimensional analysis: the derivative $\partial_y$ becomes the soft
supersymmetry-breaking parameter. Hard supersymmetry breaking effects
do not arise from the minimal kinetic terms in the 5d bulk.

So far, we have considered the gauge and Higgs fields.  How are quarks 
and leptons incorporated into the above theory?  There are essentially 
two ways to introduce quarks and leptons into the model: as fields on 
the brane or in the bulk.  Here we concentrate on the case that the 
quarks and leptons are placed on a fixed point of the $S^1/Z_2$ 
orbifold, which, without loss of generality, we take to be at $y=0$.
The case of quarks and leptons in the bulk are considered in sub-section 
\ref{subsec:mssm-bulk}.  Then, quark and lepton chiral superfields, 
$Q, U, D, L$ and $E$, are introduced on the $y=0$ brane, together with 
appropriate Yukawa couplings with the Higgs fields in the bulk,
\begin{equation}
  S = \int d^4x \; dy \; \delta(y)
	\left[ \int d^2\theta \left( y_u Q U H_1 + y_d Q D H_2^c 
	+ y_e L E H_2^c \right) + {\rm h.c.} \right].
\label{eq:yukawa}
\end{equation}
With these Yukawa couplings, the theory precisely reduces to the minimal 
supersymmetric standard model at low energies.  Note that since the 
squarks and sleptons are brane fields, their masses are not generated 
by the orbifolding; soft supersymmetry-breaking masses for squarks 
and sleptons are essentially zero at the scale of $1/R$.  However, they 
are radiatively generated through renormalization group equations below 
the scale of $1/R$.  Since the radiative corrections are almost flavor 
universal, the supersymmetric flavor problem is solved in this model.

To summarize, the present model gives the minimal supersymmetric 
standard model at low energies, with a constrained form of soft 
supersymmetry-breaking parameters.  They are given, at the scale $1/R$, by
\begin{equation}
  m_{1/2} = \hat{\alpha} \equiv \alpha/R,
\label{eq:soft-gaugino}
\end{equation}
\begin{equation}
  m_{h_u,h_d}^2 = \hat{\alpha}^2,  \qquad 
  m_{\tilde{q},\tilde{u},\tilde{d},\tilde{l},\tilde{e}}^2 = 0,  \qquad
  A = -\hat{\alpha},
\label{eq:soft-scalar}
\end{equation}
\begin{equation}
  \mu = \hat{\gamma} \equiv \gamma/R,  \qquad 
  \mu B = -2 \hat{\alpha} \hat{\gamma}.
\label{eq:soft-higgs}
\end{equation}
where $m_{1/2}$ represents the universal gaugino mass and $A$ the 
trilinear scalar couplings.  The predicted sign of $A$ is such that, 
on scaling to the infrared, $|A|$ is increased by the radiative 
contribution from the gaugino mass.  Here, we have neglected threshold 
effects coming from finite radiative corrections at $1/R$. In this 
expression, while the compactification radius $R$ is an arbitrary 
parameter, $\hat{\alpha}$ and $\hat{\gamma}$ must be around the weak 
scale for the supersymmetry to be relevant as a solution to the gauge 
hierarchy problem.  One interesting consequence of 
Eq.~(\ref{eq:soft-gaugino}) is that the gaugino masses are unified 
at the scale $1/R$. This arises because in 5d the most general 
orbifolding admits only a single parameter which breaks supersymmetry.
In general the compactification scale differs from the unification scale, 
so that the gaugino masses do not unify at the grand 
unification scale.  By construction, all the above quantities 
are necessarily real, so that there is no supersymmetric CP problem.
Below, we consider all the range of $1/R$ from the weak to the 
Planck scale, treating $\hat{\alpha}$ and $\hat{\gamma}$ as 
free parameters of the order of the weak scale.

\subsection{Radiative electroweak symmetry breaking}
\label{subsec:mssm-ewsb}

Having obtained soft supersymmetry-breaking parameters, 
Eqs.~(\ref{eq:soft-gaugino} -- \ref{eq:soft-higgs}), at the 
compactification scale, we can solve renormalization group equations 
to obtain the spectrum at the weak scale.  In particular, we can 
work out whether radiative electroweak symmetry breaking occurs 
correctly or not.  In this sub-section, we will consider the constraint 
from radiative electroweak symmetry breaking and find that it gives 
a restriction on the values for $1/R$, $\hat{\alpha}$ and 
$\hat{\gamma}$.

The minimization of the Higgs potential gives two relations,
\begin{eqnarray}
  \frac{m_Z^2}{2} &=& \frac{\tan^2\beta\, m_{h_u}^2 - m_{h_d}^2}
	{1-\tan^2\beta} - |\mu|^2, 
\label{eq:ewsb-1}\\
  \sin(2\beta) &=& -\frac{2\mu B}{m_{h_u}^2 + m_{h_d}^2 + 2|\mu|^2},
\label{eq:ewsb-2}
\end{eqnarray}
where various quantities must be evaluated at the weak scale.
Using these equations, we can relate two parameters $\hat{\alpha}$ 
and $\hat{\gamma}$ with $\tan\beta \equiv \vev{h_u}/\vev{h_d}$ and 
$v \equiv \sqrt{\vev{h_u}^2 + \vev{h_d}^2}$.  Since $v$ is fixed by 
the observed Fermi constant, there are two remaining free parameters 
which we take to be $1/R$ and $\tan\beta$.  Below, we consider the 
constraint on this two dimensional parameter space from 
electroweak symmetry breaking.  We look for solutions of 
Eqs.~(\ref{eq:ewsb-1}, \ref{eq:ewsb-2}) with $\tan\beta \gsim 2$, 
to satisfy the experimental lower bound on the physical Higgs boson mass.

A characteristic feature of the soft supersymmetry-breaking 
parameters given in Eqs.~(\ref{eq:soft-gaugino} -- \ref{eq:soft-higgs}) 
is a sizable non-vanishing value for the $\mu B$ parameter.  
The $\mu B$ parameter pushes the value of $\tan\beta$ towards $1$.
Thus we want to reduce the effect of the $\mu B$ parameter relative to 
that of the other parameters, which requires a hierarchy between the 
two parameters $\hat{\alpha}$ and $\hat{\gamma}$.
This can be seen easily as follows. Suppose, as a zero-th 
order approximation, that only $m_{h_u}^2$ is changed through 
renormalization group evolution from $1/R$ to the weak scale.
Then, relevant parameters are given by 
$m_{h_u}^2 = (1-c)\hat{\alpha}^2$, $m_{h_d}^2 = \hat{\alpha}^2$, 
$\mu = \hat{\gamma}$ and $\mu B = -2 \hat{\alpha} \hat{\gamma}$ 
at the weak scale.  Here, $c$ parameterizes the renormalization scaling 
induced by the top Yukawa coupling, and depends on $1/R$ and 
$\tan\beta$ through the distance of renormalization group running 
and the size of the top Yukawa coupling, respectively.
After this scaling, Eq.~(\ref{eq:ewsb-2}) becomes
\begin{equation}
  \frac{\tan\beta}{1 + \tan^2\beta}
    \simeq \frac{\hat{\alpha} \hat{\gamma}}
	{(1-c/2)\hat{\alpha}^2 + \hat{\gamma}^2},
\label{eq:approx-1}
\end{equation}
and $\tan\beta \gg 1$ requires either $\hat{\alpha}/\hat{\gamma} \ll 1$ 
or $\hat{\gamma}/\hat{\alpha} \ll 1$.  Although the above argument 
is very rough, numerical computations confirm that successful 
electroweak symmetry breaking with $\tan\beta \gsim 2$ requires a 
hierarchy between $\hat{\alpha}$ and $\hat{\gamma}$ of 
typically an order of magnitude.

We first consider the case $\hat{\alpha}/\hat{\gamma} \ll 1$.
In this case, electroweak symmetry breaking does not occur, since 
the supersymmetric mass term for the Higgs fields is much larger than 
the supersymmetry-breaking masses which would trigger the electroweak 
symmetry breaking.  In other words, the right-hand side of 
Eq.~(\ref{eq:ewsb-1}) formally gives a negative value and is unphysical.  
Thus we concentrate on the case $\hat{\gamma}/\hat{\alpha} \ll 1$ 
from now on.  With $\hat{\gamma}/\hat{\alpha} \ll 1$, the values 
for $\hat{\alpha}$ and $\hat{\gamma}$ are given by 
$\hat{\gamma} \ll \hat{\alpha} \sim m_Z$ in a generic region of 
the parameter space, so that it is not phenomenologically acceptable.  
This can be easily seen by noting that Eq.~(\ref{eq:ewsb-1}) reduces, 
with moderately large values for $\tan\beta$, to 
\begin{equation}
  \frac{m_Z^2}{2} \simeq -(1-c) \hat{\alpha}^2 - \hat{\gamma}^2,
\label{eq:approx-2}
\end{equation}
in the approximation adopted in Eq.~(\ref{eq:approx-1}).
However, Eq.~(\ref{eq:approx-2}) also provides the way to avoid this 
problem.  If $(c-1) \hat{\alpha}^2 \simeq \hat{\gamma}^2$, which 
means $c \simeq 1$, we can obtain $\hat{\gamma} \sim m_Z$ or even 
larger values for $\hat{\gamma}$.  Then, since $c$ depends on both 
$1/R$ and $\tan\beta$, $c \simeq 1$ gives one constraint on these 
values: a phenomenologically acceptable parameter region is a line in 
the two dimensional parameter space spanned by $1/R$ and $\tan\beta$.

The actual dependence of $c$ on $1/R$ and $\tan\beta$ is somewhat 
complicated, and also the approximation in Eq.~(\ref{eq:approx-1}) 
is not very precise.  Thus we have evaluated the allowed region by 
numerical computations, including full renormalization group effects 
at the two-loop level.  We find that the allowed region is a 
curved line, which extends from 
$(1/R, \tan\beta) \sim (2 \times 10^6~{\rm GeV}, 2)$ through 
$(1/R, \tan\beta) \sim (3 \times 10^7~{\rm GeV}, 5)$ to 
$(1/R, \tan\beta) \sim (6 \times 10^7~{\rm GeV}, 20)$.
The thickness of this line is given by $\delta (\tan\beta) /\tan\beta 
\sim 5\%$, with a fixed value of $1/R$.  This behavior is easily 
understood in terms of the correlation between $1/R$ and $\tan \beta$ 
required to maintain $c$ close to unity.  If $1/R$ is below 
$2 \times 10^6~{\rm GeV}$, the running distance is short so that 
$c \simeq 1$ requires fairly large values for the top Yukawa coupling, 
corresponding to $\tan\beta \lsim 2$.  Thus there is no phenomenologically 
acceptable parameter region for $1/R \lsim 2 \times 10^6~{\rm GeV}$.  
Once $1/R$ is increased above $2 \times 10^6~{\rm GeV}$, an allowed 
parameter region emerges, giving correct radiative electroweak symmetry 
breaking with $\tan\beta \gsim 2$.  As $1/R$ is further increased, the 
running distance also increases, and thus $c \simeq 1$ requires 
smaller values of the top Yukawa coupling, corresponding to larger 
$\tan\beta$.  Thus the allowed parameter region extends to the upper 
right direction in the $(1/R, \tan\beta)$ plane.  However, the top 
Yukawa coupling cannot be made arbitrary small, since its dependence on 
$\tan\beta$ is very weak for $\tan\beta \gsim 20$, leading to an 
upper bound: $1/R \lsim 6 \times 10^7~{\rm GeV}$.

In the region discussed above, the low energy $B$ parameter is large, 
reflecting the large value at the compactification scale.  However, 
the sign of $A$ is such that the magnitude of $B$ is reduced during 
evolution to the infrared; hence for large enough $1/R$ we find an 
acceptable region with a low value for $B$ at the weak scale.
This requires $1/R \gsim 10^{14}~{\rm GeV}$, in which case we find 
successful electroweak symmetry breaking occurs only for $\tan\beta$ 
near $2$.  Since $\hat{\gamma}/\hat{\alpha}$ is now of order unity, 
there is no need for $c \simeq 1$.  However, for $\tan\beta \simeq 2$, 
a sufficiently heavy Higgs boson only results for heavy squarks, 
so $\hat{\alpha}$ must be large.  Hence some cancellation between 
the $\hat{\alpha}^2$ and $\hat{\gamma}^2$ terms are required in 
Eq.~(\ref{eq:approx-2}).

In summary, we have found two regions of parameter space where 
correct electroweak symmetry breaking occurs in the present model.
In one region, the compactification scale is tightly constrained, 
$2 \times 10^6~{\rm GeV} \lsim 1/R \lsim 6 \times 10^7~{\rm GeV}$, while 
a broad region of $\tan\beta$, $2 \lsim \tan\beta \lsim 20$, can be 
realized depending on the value of $1/R$.  This somewhat unusual 
result is a consequence of the sizable $\mu B$ parameter at the 
compactification scale.  In the other region, $1/R \gsim 
10^{14}~{\rm GeV}$ and $\tan\beta \simeq 2$.

The above constraint on $1/R$ and $\tan\beta$ is derived by considering 
only the usual logarithmic renormalization group evolutions.  
There are also finite one-loop radiative corrections to the soft 
supersymmetry-breaking parameters that do not involve a log factor.  
In a 5d calculation, these appear as threshold effects at $1/R$. 
In the 4d picture this corresponds to including supersymmetry 
breaking effects from higher KK modes. These contributions are 
expected to be of $O(\hat{\alpha}^2/16\pi^2)$ and thus smaller than 
those calculated above by an amount of order $1/ \ln(1 / \alpha)$.  
To find whether they are negligible or significant, however, a full 
one-loop calculation must be done to include the effects of the heavier 
modes of the KK tower so that a finite answer is obtained. In the case 
that the usual logarithmic term dominates, the allowed range of $1/R$ 
quoted above will be unchanged. 

Finally, we comment on the effect of brane-localized kinetic terms 
for the Higgs fields,
\begin{equation}
  S = \int d^4x \; dy \; \delta(y)
	\int d^4\theta \left( Z_{u} H_u^\dagger H_u 
	+ Z_{d} H_d^\dagger H_d \right).
\label{eq:brane-kin}
\end{equation}
This changes the Higgs potential and Higgsino mass given in 
Eq.~(\ref{eq:higgs-pot}) to
\begin{eqnarray}
  {\cal L} &=& 
	- \left( \frac{\alpha_u^{\prime 2}}{R^2} + 
	\frac{\gamma^{\prime 2}}{R^2} \right) |h_u|^2
	- \left( \frac{\alpha_d^{\prime 2}}{R^2} + 
	\frac{\gamma^{\prime 2}}{R^2} \right) |h_d|^2 \nonumber\\
  && + \frac{(\alpha'_u+\alpha'_d)\gamma'}{R^2} (h_u h_d + {\rm h.c.})
	- \frac{\gamma'}{R} (\tilde{h}_u \tilde{h}_d + {\rm h.c.}).
\end{eqnarray}
Here, $\alpha'_u, \alpha'_d$ and $\gamma'$ are given by
\begin{equation}
  \alpha'_u = \frac{\alpha}{1+z_u}, \qquad
  \alpha'_d = \frac{\alpha}{1+z_d}, \qquad
  \gamma' = \frac{\gamma}{\sqrt{1+z_u}\sqrt{1+z_d}},
\end{equation}
where $z_u \equiv Z_u / 2\pi R$ and $z_d \equiv Z_d / 2\pi R$.
However, we expect that $z_u$ and $z_d$ are small, since they are 
suppressed by the length of the extra dimension.  We have checked 
that the inclusion of the brane kinetic terms does not change 
the qualitative feature of the analysis in this sub-section unless 
they are large, $z_u, z_d \gsim O(1)$.

\subsection{Quarks and leptons in the bulk}
\label{subsec:mssm-bulk}

In this sub-section, we consider the case where the quarks and leptons 
are put in the bulk, rather than on the $y=0$ brane.
In this case, we have to introduce hypermultiplets ${\cal Q}_j, 
{\cal U}_j, {\cal D}_j, {\cal L}_j$ and ${\cal E}_j$ in the 5d bulk, 
where $j=1,2,3$ represents the generation index.  Each of them is 
decomposed, under 4d $N=1$ supersymmetry, into two chiral superfields 
as $(Q_j, Q_j^c), (U_j, U_j^c), (D_j, D_j^c), (L_j, L_j^c)$ and 
$(E_j, E_j^c)$ where conjugated fields have conjugate transformations 
under the gauge group.

The boundary conditions under the orbifolding are given as follows.
We must require that the orbifolding yields three light generations of
chiral matter. This uniquely determines that, under $y \leftrightarrow
-y$, ${\cal Q}_j$'s obey 
\begin{equation}
  \pmatrix{Q_1 & Q_2 & Q_3 \cr 
	Q_1^{c\dagger} & Q_2^{c\dagger} & Q_3^{c\dagger}}(x^\mu,-y) 
  = \pmatrix{Q_1 & Q_2 & Q_3 \cr 
	-Q_1^{c\dagger} & -Q_2^{c\dagger} & -Q_3^{c\dagger}}(x^\mu,y).
\end{equation}
and also that, under $y \leftrightarrow y+2\pi R$
\begin{eqnarray}
  \pmatrix{\tilde{q}_1 & \tilde{q}_2 & \tilde{q}_3 \cr 
	\tilde{q}_1^{c\dagger} & \tilde{q}_2^{c\dagger} & 
	\tilde{q}_3^{c\dagger}}(x^\mu,y+2\pi R) 
	&=& e^{-2\pi i \alpha \sigma_2} 
	\pmatrix{\tilde{q}_1 & \tilde{q}_2 & \tilde{q}_3 \cr 
	\tilde{q}_1^{c\dagger} & \tilde{q}_2^{c\dagger} & 
	\tilde{q}_3^{c\dagger}}(x^\mu,y), 
\\
  \pmatrix{q_1 & q_2 & q_3 \cr 
	q_1^{c\dagger} & q_2^{c\dagger} & q_3^{c\dagger}}(x^\mu,y+2\pi R) 
	&=& \pmatrix{q_1 & q_2 & q_3 \cr 
	q_1^{c\dagger} & q_2^{c\dagger} & q_3^{c\dagger}}(x^\mu,y),
\end{eqnarray}
where $Q_j = (\tilde{q}_j, q_j)$ and $Q^c_j = (\tilde{q}^c_j, q^c_j)$.
The uniqueness of this choice is non-trivial. A twisting of the fields
in flavor space under $y \leftrightarrow y+2\pi R$ is only consistent 
if the fields have opposite parities under $y \leftrightarrow -y$ 
\cite{Barbieri:2001dm}. However, having opposite parities yields 
vector-like matter at low energy, as can be seen from the Higgs case.
The other fields, ${\cal U}_j, {\cal D}_j, {\cal L}_j$ and ${\cal E}_j$, 
must obey the same boundary conditions.  With these boundary conditions, 
the matter content below $1/R$ scale is exactly the three generations 
of quark and lepton chiral superfields.

The mode expansions for the squarks are given by
\begin{equation}
  \pmatrix{\tilde{q}_j \cr \tilde{q}_j^{c\dagger}}(x^\mu,y) 
	= \sum_{n=0}^{\infty} e^{-i \alpha \sigma_2 y / R} 
	\pmatrix{\tilde{q}_{jn} \cos[ny/R] \cr 
	\tilde{q}_{jn}^{c\dagger} \sin[ny/R]}.
\label{eq:squarkmodes}
\end{equation}
The expansions for the quarks are straightforward (corresponding to 
$\alpha=0$ in the above equation).  Identifying $n=0$ modes with the 
usual squarks (and sleptons), we obtain the following soft 
supersymmetry-breaking masses at the $1/R$ scale:
\begin{equation}
  {\cal L} = -\frac{\alpha^2}{R^2} \sum_{j=1}^{3}
	\left( |\tilde{q}_j|^2 + |\tilde{u}_j|^2 + |\tilde{d}_j|^2
	+ |\tilde{l}_j|^2 + |\tilde{e}_j|^2 \right),
\end{equation}
where squark and slepton fields are canonically normalized in 4d.
We find that the universal scalar mass, $\alpha/R$, is generated.
This degeneracy among various squark and slepton masses is lifted by 
the presence of the brane-localized kinetic terms.  However, we expect 
that these terms are small due to the volume suppression from the 
extra dimension, so that these theories offer an interesting way 
to solve the supersymmetric flavor problem.

As in the case of brane matter, the Yukawa couplings 
Eq.~(\ref{eq:yukawa}) are introduced on the $y=0$ brane.
Then, below the compactification scale $1/R$, the theory reduces 
to the minimal supersymmetric standard model with 
\begin{equation}
  m_{1/2} = \hat{\alpha},  \qquad
  m_{h_u,h_d}^2 = 
  m_{\tilde{q},\tilde{u},\tilde{d},\tilde{l},\tilde{e}}^2 = 
  \hat{\alpha}^2,  \qquad 
  A = -3 \hat{\alpha},
\label{eq:soft-1}
\end{equation}
\begin{equation}
  \mu = \hat{\gamma},  \qquad 
  \mu B = -2 \hat{\alpha} \hat{\gamma},
\label{eq:soft-2}
\end{equation}
Again, there are neither grand unified relations among the gaugino 
masses (unless $1/R$ is close to the unification scale) nor 
supersymmetric flavor or CP problems.  

In contrast to the case of brane matter, squarks and sleptons have 
non-vanishing soft masses at the compactification scale and the $A$ 
parameter is very large.  This substantially changes the numerical 
results for successful electroweak symmetry breaking, although the 
qualitative behavior is the same: the allowed parameter region is a 
line in the $(1/R, \tan\beta)$ plane with the two quantities positively 
correlated. There are both a low $1/R$ region, having $c$ close to unity 
and large $B$, and a high $1/R$ region, with $c$ not near unity and 
smaller values for $B$.  In the low $1/R$ region, the values for $1/R$ 
are much lower than those in the brane matter case, since the 
non-vanishing top squark masses at the compactification scale and the 
large $A$ parameter lead to large radiative corrections for the Higgs 
mass, so that $c \simeq 1$ is obtained with a shorter running distance. 
Successful electroweak symmetry breaking occurs in this region with 
$700~{\rm GeV} \lsim 1/R \lsim 2~{\rm TeV}$, with $2 \lsim \tan\beta 
\lsim 20$. The second region is larger than before, since small $B$ 
is obtained with less running due to the larger value for $A$. This 
region extends from $(1/R, \tan\beta) \sim (10^7~{\rm GeV}, 2)$ to 
$(1/R, \tan\beta) \sim (10^{16}~{\rm GeV}, 4)$.  We conclude that there 
is now a preferred region: $1/R$ can be identified with the unification 
scale, and the resulting value for $\tan\beta$ is sufficiently large 
that the Higgs mass bound does not require the squarks to be very heavy, 
so that electroweak symmetry breaking occurs with little fine tuning. 
Finally we note that there is a small third region having small $B$ 
giving $(1/R, \tan\beta) \sim (10^{16}~{\rm GeV}, 30)$, so that the 
$b$-quark Yukawa coupling is sufficiently large to affect the scaling 
of the Higgs mass parameters.

\subsection{Alternative sources for $\mu$}
\label{subsec:mu}

So far in this paper we have assumed that both supersymmetry breaking 
and $G_H$ breaking arise from boundary conditions at the scale $1/R$. 
Here we comment briefly on alternative sources for $\mu$, while 
preserving supersymmetry breaking from the boundary condition parameter 
$\alpha$. If $\mu$ arises from physics at shorter distances than $1/R$, 
then it will appear in the five dimensional theory as a brane 
localized operator $\delta(y) \mu H_1 H_2^c$. In this case the 
compactification leads to $B=-2\hat{\alpha}$ at the scale $1/R$, so
that the regions for successful electroweak symmetry breaking are 
precisely those discussed in the previous two sub-sections.  However, 
if $\mu$ is generated in the four dimensional effective theory below 
$1/R$, then other values of $B$ will occur in general, changing the 
conditions for electroweak symmetry breaking.

As an example of low scale $\mu$ generation, we consider a theory
which is similar to the next-to-minimal supersymmetric 
standard model.  We introduce a singlet chiral superfield $S$ on the 
$y=0$ brane and couple it with the Higgs fields as
\begin{equation}
  S = \int d^4x \; dy \; \delta(y)
	\left[ \int d^2\theta \left( \lambda\, S H_1 H_2^c 
	+ \frac{k}{3}\, S^3 \right) 
	+ {\rm h.c.} \right].
\label{eq:nmssm}
\end{equation}
A tree level $\mu$ parameter may be forbidden by an $R$ or discrete 
$Z_3$ symmetry, or by the requirement that the superpotential not 
contain any mass parameter.  At the compactification scale, 
$A_\lambda = -2\hat{\alpha}$ and $A_k = m_s^2 = 0$.  The Higgs 
fields have soft supersymmetry-breaking masses at the compactification 
scale, so that renormalization group scaling below $1/R$ drives 
$m_s^2$ negative.  The resulting vacuum expectation values for the 
scalar and $F$ components of $S$ generate effective $\mu$ and $\mu B$ 
parameters, respectively.  Since these expectation values depend on 
the coupling constants $\lambda$ and $k$, the $\mu$ and $\mu B$ 
parameters are essentially free parameters in this model.  Therefore, 
we can evade the stringent constraints on $(1/R, \tan\beta)$ derived 
in the previous sub-sections.  However, supersymmetry breaking is 
still determined by the single parameter $\hat{\alpha}$, so that the 
tight predictions for squark, slepton and gaugino masses still apply.

\section{Embedding into SU(5)}
\label{sec:su5}

In this section, we construct 5d $SU(5)$ theories which reduce to the 
softly broken minimal supersymmetric standard model at low energies.
The structure of the theories is similar to the 5d $SU(5)$ model discussed 
in Refs.~\cite{Kawamura:2000ev, Hall:2001pg}.  However, the orbifold 
boundary conditions are modified using $U(1)_R$ and $U(1)_H$, giving 
simultaneous breakings of both supersymmetry and $SU(5)$ gauge symmetry 
from a single orbifolding.

\subsection{The model}

We consider a 5d $SU(5)$ gauge theory compactified on the $S^1/Z_2$ 
orbifold.  The radius of the fifth dimension is taken to be around 
the grand unification scale, $1/R \sim 10^{16}~{\rm GeV}$, as we 
will see later.  We also introduce two Higgs hypermultiplets, 
${\cal H}_{\bf 5_1} = (H_{\bf 5_1}, H_{\bf 5_1}^{c\dagger})$ and 
${\cal H}_{\bf 5_2} = (H_{\bf 5_2}, H_{\bf 5_2}^{c\dagger})$, in the 
bulk, each transforming as a fundamental representation of $SU(5)$.

What accomplishes the $SU(5)$ breaking?  We impose that the gauge 
and Higgs fields transform as ${\cal V} \rightarrow P {\cal V} P^{-1}$ 
and ${\cal H}_{\bf 5_{1,2}} \rightarrow P {\cal H}_{\bf 5_{1,2}}$
under $y \rightarrow y+2\pi R$.  Here, $P$ is a diagonal matrix acting on 
the index of the fundamental representation, $P = {\rm diag}(-,-,-,+,+)$.
This gives masses of order $1/R$ for the gauge bosons of 
$SU(5) / (SU(3) \times SU(2) \times U(1))$ and for the triplet Higgs 
fields, so that the effective field theory below $1/R$ is that of 
a 4d, $N=1$, $SU(3) \times SU(2) \times U(1)$ gauge theory with two 
Higgs doublets.  An important point here is that this boundary conditions 
are compatible with those discussed in the previous section 
which were used to break supersymmetry and give the $\mu$ term.  
That is, we can simultaneously impose both $SU(5)$ breaking and 
supersymmetry breaking boundary conditions.

To show how the above construction works explicitly, let us 
label the gauge fields of $SU(3) \times SU(2) \times U(1)$ 
and $SU(5) / (SU(3) \times SU(2) \times U(1))$ as 
${\cal V}^{(+)} = (V^{(+)}, \Sigma^{(+)})$ and 
${\cal V}^{(-)} = (V^{(-)}, \Sigma^{(-)})$, respectively.
Similarly, we represent the doublet and triplet components of the 
Higgs hypermultiplets by the superscript $(+)$ and $(-)$, 
respectively: ${\cal H}_{\bf 5_i} \rightarrow {\cal H}_i^{(\pm)} = 
(H_i^{(\pm)}, H_i^{(\pm)c\dagger})$ where $i = 1,2$.
Then, the boundary conditions are explicitly represented as follows.
Under $y \leftrightarrow -y$, the fields must satisfy 
\begin{eqnarray}
  \pmatrix{V^{(\pm)} \cr \Sigma^{(\pm)}}(x^\mu,-y) 
  &=& \pmatrix{V^{(\pm)} \cr -\Sigma^{(\pm)}}(x^\mu,y), \\
  \pmatrix{H_1^{(\pm)} & H_2^{(\pm)} \cr 
	H_1^{(\pm)c\dagger} & H_2^{(\pm)c\dagger}}(x^\mu,-y) 
  &=& \pmatrix{H_1^{(\pm)} & -H_2^{(\pm)} \cr 
	-H_1^{(\pm)c\dagger} & H_2^{(\pm)c\dagger}}(x^\mu,y),
\end{eqnarray}
and, under $y \leftrightarrow y+2\pi R$, they obey 
\begin{eqnarray}
  A^{(\pm)M}(x^\mu,y+2\pi R) &=& \pm\, A^{(\pm)M}(x^\mu,y), 
\\
  \pmatrix{\lambda^{(\pm)} \cr \lambda^{\prime(\pm)}}(x^\mu,y+2\pi R) &=& 
	\pm\, e^{-2\pi i \alpha \sigma_2} 
	\pmatrix{\lambda^{(\pm)} \cr \lambda^{\prime(\pm)}}(x^\mu,y), 
\\
  \sigma^{(\pm)}(x^\mu,y+2\pi R) &=& \pm\, \sigma^{(\pm)}(x^\mu,y), 
\\ \nonumber\\
  \pmatrix{h_1^{(\pm)} & h_2^{(\pm)} \cr 
	h_1^{(\pm)c\dagger} & h_2^{(\pm)c\dagger}}(x^\mu,y+2\pi R) 
	&=& \pm\, e^{-2\pi i \alpha \sigma_2} 
	\pmatrix{h_1^{(\pm)} & h_2^{(\pm)} \cr 
	h_1^{(\pm)c\dagger} & h_2^{(\pm)c\dagger}}
	e^{2\pi i \gamma \sigma_2} (x^\mu,y), 
\\
  \pmatrix{\tilde{h}_1^{(\pm)} & \tilde{h}_2^{(\pm)} \cr 
	\tilde{h}_1^{(\pm)c\dagger} & 
	\tilde{h}_2^{(\pm)c\dagger}}(x^\mu,y+2\pi R) 
	&=& \pm \pmatrix{\tilde{h}_1^{(\pm)} & \tilde{h}_2^{(\pm)} \cr 
	\tilde{h}_1^{(\pm)c\dagger} & \tilde{h}_2^{(\pm)c\dagger}}
	e^{2\pi i \gamma \sigma_2} (x^\mu,y).
\end{eqnarray}
Here, we are considering $\alpha \sim \gamma \ll 1$ such that 
$\alpha/R \sim \gamma/R$ are around the weak scale.

In the limit $\alpha, \gamma \rightarrow 0$, the above boundary 
conditions give the following mass spectrum \cite{Kawamura:2000ev}.
The fields in the minimal supersymmetric standard model, $V^{(+)}, 
H_1^{(+)}$ and $H_2^{(+)c}$, have a tower with masses given by 
$n/R$ ($n = 0,1,2,\cdots$); similarly, $(n+1/2)/R$ for $V^{(-)}, 
\Sigma^{(-)}, H_1^{(-)}, H_1^{(-)c}, H_2^{(-)}$ and $H_2^{(-)c}$, 
and $(n+1)/R$ for $\Sigma^{(+)}, H_1^{(+)c}$ and $H_2^{(+)}$.
Therefore, in this limit, we have massless fields, $V^{(+)}, 
H_1^{(+)} \equiv H_u$ and $H_2^{(+)c} \equiv H_d$, which we call 
quasi zero-modes.  Furthermore, since the broken gauge fields 
have masses of $O(1/R)$, the compactification scale is of order 
the unification scale.

When we turn on tiny non-zero values for $\alpha \sim \gamma$, 
they perturb the mass spectrum of the towers by an amount 
$O(\alpha/R \sim \gamma/R)$.  In particular, it gives the soft 
supersymmetry-breaking masses for the quasi zero-modes and the $\mu$ 
term.  Since the boundary conditions for the quasi zero-modes are 
the same as those discussed in section \ref{sec:mssm}, the effective 
4d Lagrangian for the supersymmetry and $G_H$ breaking interactions 
below $1/R$ is given by Eq.~(\ref{eq:higgs-pot}).  This forces us to 
take $\alpha/R \sim \gamma/R$ around the weak scale, that is 
$\alpha \sim \gamma \sim 10^{-13}$.

Before introducing quarks and leptons into the model, let us note one 
important difference between the $SU(3) \times SU(2) \times U(1)$ model
and the present $SU(5)$ model.  In the $SU(3) \times SU(2) \times U(1)$ 
case, the gaugino masses are unified at the compactification scale, 
so that there is generically no grand unified relation among them.
On the other hand, in the $SU(5)$ case, the grand unified 
relation $m_{\lambda_{SU(3)}}/\alpha_{SU(3)} = 
m_{\lambda_{SU(2)}}/\alpha_{SU(2)} = m_{\lambda_{U(1)}}/\alpha_{U(1)}$ 
necessarily holds.  This is true even in the presence of 
$SU(5)$-violating gauge kinetic terms that can be introduced on 
the $y=\pi R$ brane.  The argument is the following.
Suppose we have both the bulk gauge kinetic term, which must be 
$SU(5)$ symmetric, and the brane-localized gauge kinetic terms at 
$y=\pi R$, which can have different coefficients for $SU(3)$, 
$SU(2)$ and $U(1)$.  Then, the 4d gauge couplings $g_a$ are given by 
$1/g_a^2 = 2\pi R/g_5^2 + 1/g_{4,a}^2$, where $g_5$ and $g_{4,a}^2$ 
are the bulk and brane gauge couplings, respectively, and $a$ runs 
over $SU(3)$, $SU(2)$ and $U(1)$.  Thus these gauge couplings are not 
necessarily unified exactly at the compactification scale, although the 
$SU(5)$-violating piece is volume suppressed and small \cite{Hall:2001pg}. 
On the other hand, the gaugino masses $m_{\lambda,a}$ are given by 
$m_{\lambda,a}/g_a^2 = (2\pi R/g_5^2)(\alpha/R)$.  This shows that 
the quantities $m_{\lambda,a}/g_a^2$ are universal, and thus the grand 
unified relation on the gaugino masses holds very precisely.

Let us now discuss the quarks and leptons.  As in the case of 
$SU(3) \times SU(2) \times U(1)$, we can introduce them either 
on the brane or in the bulk.  We first consider the case of brane 
matter, and defer the case of bulk matter to the next sub-section.
To obtain the usual $SU(5)$ understanding of quark and lepton quantum
numbers, we introduce matter chiral superfields $T_{\bf 10_j}$ and 
$F_{\bar{\bf 5}_{\bf j}}$ on the $y=0$ brane, where $j=1,2,3$ is 
the generation index.  Then, we can write down the $SU(5)$ symmetric 
Yukawa couplings on the $y=0$ brane 
\cite{Kawamura:2000ev, Hall:2001pg}\footnote{
These brane interactions are different from those adopted in 
Ref.~\cite{Altarelli:2001qj}.}
\begin{equation}
  S = \int d^4x \; dy \; \delta(y)
	\left[ \int d^2\theta \sum_{j,k=1}^{3} \left( 
	(y_1)_{jk} T_{\bf 10_j} T_{\bf 10_k} H_{\bf 5_1} + 
	(y_2)_{jk} T_{\bf 10_j} F_{\bar{\bf 5}_{\bf k}} H_{\bf 5_2}^c 
	\right) + {\rm h.c.} \right].
\label{eq:su5-yukawa}
\end{equation}
With these Yukawa couplings, the theory reduces, below the 
compactification scale, to the minimal supersymmetric standard model 
with the soft supersymmetry-breaking parameters (and the $\mu$ parameter) 
given by Eqs.~(\ref{eq:soft-gaugino} -- \ref{eq:soft-higgs}).
Here electroweak symmetry breaking is as before except now we require 
$1/R \approx 10^{16}~{\rm GeV}$ so that $\tan\beta \simeq 2$.
The Higgs mass bound is only satisfied for somewhat heavy squarks, 
giving some fine tuning in electroweak symmetry breaking.

We finally comment on the phenomenologies of the present model.
First of all, the dangerous dimension 5 proton decay operators 
are not generated by an exchange of the triplet Higgs multiplets,
due to the specific form of the triplet Higgs mass terms 
\cite{Hall:2001pg}.  Furthermore, unwanted tree-level brane operators 
at $y=0$, such as $[H_{\bf 5_1} H_{\bf 5_2}^c]_{\theta^2}, 
[T_{\bf 10_j} T_{\bf 10_k} T_{\bf 10_l} F_{\bar{\bf 5}_{\bf m}}]_{\theta^2}, 
[F_{\bar{\bf 5}_{\bf j}} H_{\bf 5_1}]_{\theta^2}$ and 
$[T_{\bf 10_j} F_{\bar{\bf 5}_{\bf k}} F_{\bar{\bf 5}_{\bf l}}]_{\theta^2}$, 
are forbidden by imposing $U(1)_R$ symmetry on the theory 
\cite{Hall:2001pg}. (In the case of the non-minimal theory, with 
the superpotential of Eq.~(\ref{eq:nmssm}), the $U(1)_R$ symmetry 
given in Ref.~\cite{Hall:2001pg} is explicitly broken to a $Z_{4,R}$ 
subgroup, which is still sufficient to forbid these unwanted operators. 
An alternative way is to consider a different U(1)$_R$ symmetry, under 
which $\{ H_{\bf 5_1}, H_{\bf 5_2}^c, T_{\bf 10_j}, 
F_{\bar{\bf 5}_{\bf k}}, S \}$ and $\{ H_{\bf 5_1}^c, 
H_{\bf 5_2} \}$ have charges $2/3$ and $4/3$, respectively.)
This $U(1)_R$ symmetry (or $Z_{4,R}$) is weakly broken to 
$R$-parity subgroup by the orbifold boundary conditions, so that 
the gaugino masses and the $\mu$ parameter are generated.
The value of $1/R$ is lower than the conventional grand unification 
scale $\simeq 2 \times 10^{16}~{\rm GeV}$ due to the threshold effect 
coming from the KK towers \cite{Hall:2001pg, Nomura:2001mf}.
It predicts a higher rate for the dimension 6 proton decay 
than in the usual 4d grand unified theories, which might be seen by 
further running of the Super-Kamiokande experiment or at a next 
generation proton decay detector \cite{Hall:2001pg}.  The soft 
supersymmetry-breaking masses for the squarks and sleptons are 
vanishing at the compactification scale, providing a solution to 
the supersymmetric flavor problem.

\subsection{Matter in the bulk}

In this sub-section, we introduce matter in the 5d bulk 
instead of on the $y=0$ brane.  One might naively think 
that we have only to introduce hypermultiplets ${\cal T}_{\bf 10}$ 
and ${\cal F}_{\bar{\bf 5}}$ for each generation, to obtain 
the correct low energy matter content.  However, this does not work 
because of an automatic ``double-triplet splitting'' mechanism operating 
in this setup.  Let us, for example, consider ${\cal F}_{\bar{\bf 5}} = 
(F_{\bar{\bf 5}}(+), F_{\bar{\bf 5}}^c (-))$ transforming as 
${\cal F}_{\bar{\bf 5}} \rightarrow {\cal F}_{\bar{\bf 5}} P^{-1}$ 
under $y \rightarrow y+2\pi R$, where the signs in the 
parentheses represent the parities under $y \rightarrow -y$. 
After the orbifolding, this hypermultiplet gives only one quasi
zero-mode, which is the lepton doublet $L$ of the standard model. 
Thus we do not obtain the correct matter content, $D$ and $L$.
To evade this problem, we can introduce another 
hypermultiplet ${\cal F}'_{\bar{\bf 5}} = 
(F'_{\bar{\bf 5}}(+), F_{\bar{\bf 5}}^{\prime c} (-))$ which transforms 
as ${\cal F}'_{\bar{\bf 5}} \rightarrow -{\cal F}'_{\bar{\bf 5}} P^{-1}$
under $y \rightarrow y+2\pi R$.  This additional hypermultiplet gives 
$D$ of the standard model as a quasi zero-mode, and completes the 
standard model matter content.  A similar argument applies to 
the ${\cal T}_{\bf 10}$ multiplet: the quasi zero-modes from 
${\cal T}_{\bf 10}$ are $U,E$, while that from ${\cal T}'_{\bf 10}$ 
is $Q$.  Therefore, to ensure the correct low energy matter 
content, we introduce four hypermultiplets, 
${\cal T}_{\bf 10}, {\cal T}'_{\bf 10}, {\cal F}_{\bar{\bf 5}}$ 
and ${\cal F}'_{\bar{\bf 5}}$, for each generation \cite{Hall:2001pg}.

We here explicitly show the boundary conditions for the bulk matter 
fields in the present model.  To do so, we introduce the following 
notation.  We represent $U,E/Q$ components of ${\cal T}_{\bf 10}$ 
(and corresponding states of ${\cal T}'_{\bf 10}$) by the superscripts 
$(+)/(-)$, respectively.  We also denote $L$ and $D$ components of 
${\cal F}_{\bar{\bf 5}}$ (and corresponding states of 
${\cal F}'_{\bar{\bf 5}}$) by $(+)$ and $(-)$ superscripts.  
Then, under $y \leftrightarrow -y$, they subject to
\begin{eqnarray}
  \pmatrix{T^{(\pm)}_{\bf 10_j} \cr 
	T_{\bf 10_j}^{c (\pm) \dagger}}(x^\mu,-y) 
  &=& \pmatrix{T^{(\pm)}_{\bf 10_j} \cr 
	-T_{\bf 10_j}^{c (\pm) \dagger}}(x^\mu,y),
\\
  \pmatrix{T^{\prime (\pm)}_{\bf 10_j} \cr 
	T_{\bf 10_j}^{\prime c (\pm) \dagger}}(x^\mu,-y) 
  &=& \pmatrix{T^{\prime (\pm)}_{\bf 10_j} \cr 
	-T_{\bf 10_j}^{\prime c (\pm) \dagger}}(x^\mu,y),
\end{eqnarray}
where $j=1,2,3$ represents the generation index.
The boundary condition under $y \leftrightarrow y+2\pi R$ is given by
\begin{eqnarray}
  \pmatrix{\phi^{(\pm)}_{\bf 10_j} \cr 
	\phi_{\bf 10_j}^{c (\pm) \dagger}}(x^\mu,y + 2\pi R) 
  &=& \pm\, e^{-2\pi i \alpha \sigma_2}
	\pmatrix{\phi^{(\pm)}_{\bf 10_j} \cr 
	\phi_{\bf 10_j}^{c (\pm) \dagger}}(x^\mu,y),
\\
  \pmatrix{\phi^{\prime (\pm)}_{\bf 10_j} \cr 
	\phi_{\bf 10_j}^{\prime c (\pm) \dagger}}(x^\mu,y + 2\pi R) 
  &=& \mp\, e^{-2\pi i \alpha \sigma_2}
	\pmatrix{\phi^{\prime (\pm)}_{\bf 10_j} \cr 
	\phi_{\bf 10_j}^{\prime c (\pm) \dagger}}(x^\mu,y),
\\ \nonumber\\
  \pmatrix{\psi^{(\pm)}_{\bf 10_j} \cr 
	\psi_{\bf 10_j}^{c (\pm) \dagger}}(x^\mu,y + 2\pi R) 
  &=& \pm \pmatrix{\psi^{(\pm)}_{\bf 10_j} \cr 
	\psi_{\bf 10_j}^{c (\pm) \dagger}}(x^\mu,y),
\\
  \pmatrix{\psi^{\prime (\pm)}_{\bf 10_j} \cr 
	\psi_{\bf 10_j}^{\prime c (\pm) \dagger}}(x^\mu,y + 2\pi R) 
  &=& \mp \pmatrix{\psi^{\prime (\pm)}_{\bf 10_j} \cr 
	\psi_{\bf 10_j}^{\prime c (\pm) \dagger}}(x^\mu,y),
\end{eqnarray}
where $\phi^{(\pm)}_{\bf 10_j}$ and $\psi^{(\pm)}_{\bf 10_j}$ 
($\psi^{\prime (\pm)}_{\bf 10_j}$ and $\psi^{\prime (\pm)}_{\bf 10_j}$) 
are the scalar and fermion components of $T^{(\pm)}_{\bf 10_j}$ 
($T^{\prime (\pm)}_{\bf 10_j}$), respectively.
The same boundary condition also applies to ${\cal F}_{\bar{\bf 5}}$ 
and ${\cal F}'_{\bar{\bf 5}}$ fields.

The above boundary conditions precisely give the quarks and leptons 
in the minimal supersymmetric standard model.  Together with the 
Yukawa couplings 
\begin{eqnarray}
  S &=& \int d^4x \; dy \; \delta(y)
	\Biggl[ \int d^2\theta \sum_{j,k=1}^{3} \Bigl( 
	(y_1^1)_{jk} T_{\bf 10_j} T_{\bf 10_k} H_{\bf 5_1} + 
	(y_1^2)_{jk} T_{\bf 10_j} T'_{\bf 10_k} H_{\bf 5_1} + 
	(y_1^3)_{jk} T'_{\bf 10_j} T'_{\bf 10_k} H_{\bf 5_1} 
\nonumber\\
&& +    (y_2^1)_{jk} T_{\bf 10_j} F_{\bar{\bf 5}_{\bf k}} H_{\bf 5_2}^c +
	(y_2^2)_{jk} T_{\bf 10_j} F'_{\bar{\bf 5}_{\bf k}} H_{\bf 5_2}^c +
	(y_2^3)_{jk} T'_{\bf 10_j} F_{\bar{\bf 5}_{\bf k}} H_{\bf 5_2}^c +
	(y_2^4)_{jk} T'_{\bf 10_j} F'_{\bar{\bf 5}_{\bf k}} H_{\bf 5_2}^c
	\Bigr) + {\rm h.c.} \Biggr],
\label{eq:su5-yukawa-bulk}
\end{eqnarray}
the theory reduces to the minimal supersymmetric standard model 
at low energies.  The soft supersymmetry-breaking parameters 
(and the $\mu$ parameter) at the compactification ($\simeq$ unification) 
scale are given by Eqs.~(\ref{eq:soft-1}, \ref{eq:soft-2}).
Here electroweak symmetry breaking can occur naturally in this theory 
with $\tan\beta \simeq 4$, without the need to make the top squarks 
very heavy.

We finally discuss the phenomenologies of the model with matters 
in the bulk.  In this case, the quarks and leptons which would be 
unified into a single multiplet in the usual 4d grand unified theories 
come from different $SU(5)$ multiplets.  Specifically, $D$ and $L$ 
($Q$ and $U,E$) come from different (hyper)multiplets.
Therefore, proton decay from broken gauge boson exchange is absent
at leading order \cite{Hall:2001pg}.
Furthermore, there is no unwanted $SU(5)$ relation among the 
low energy Yukawa couplings arising from the interactions given in 
Eq.~(\ref{eq:su5-yukawa-bulk}) \cite{Hall:2001pg}.
This is reminiscent of the situation in certain string motivated
theories \cite{Witten:1985xc}.  Nevertheless, the theory still 
keeps the desired features of the usual 4d grand unified theory: 
the quantization of hypercharge and the unification of the 
three gauge couplings \cite{Hall:2001pg}.  Therefore, this type of 
theory, with matter in the bulk, preserves (experimentally) desired 
features of 4d grand unified theories, while not necessarily having the 
problematic features, such as proton decay and fermion mass relations.

\section{Conclusions}

In this paper we have introduced a new implementation of the boundary 
condition supersymmetry breaking mechanism, which allows for a 
large energy desert in which physics is described by a 4d effective
theory with softly broken supersymmetry, such as the minimal 
supersymmetric standard model. This is accomplished by having a 
boundary condition which mixes components of superfields in a 
supersymmetry breaking way by an extremely small angle, $\alpha$.

In general, we are interested in a higher dimensional supersymmetric
field theory which leads to two Higgs doublet zero-modes where there
is an orbifold symmetry with the group element
\begin{equation}
  U = e^{2\pi i \alpha T}; \qquad T = T_R + r T_H.
\label{eq:orbifoldsym}
\end{equation}
under $y \rightarrow y + 2\pi R$.
Here, $T_R$ is a generator which acts non-trivially within a 
supermultiplet of the higher dimensional theory, and $T_H$ is a 
generator which mixes up the two Higgs supermultiplets. The parameter 
$r$ is of order unity, so that the generator $T$ of the orbifold 
symmetry is a generic linear combination of $T_R$ and $T_H$. In 5d, 
there is a unique choice for these generators, and hence a unique result 
for the form of the supersymmetry breaking and $G_H$ breaking operators 
that result in the low energy 4d effective theory, as shown in 
Eq.~(\ref{eq:higgs-pot}), where $\gamma = r \alpha$.  This single 
orbifold symmetry provides a unified origin for both soft
supersymmetry-breaking parameters and the $\mu$ parameter.  The result 
of Eq.~(\ref{eq:higgs-pot}) is rather robust and relies on there being 
two light Higgs doublets, as required for gauge coupling unification.
It does not change if there are other heavy Higgs hypermultiplets
which are mixed with the light ones at order $\alpha$ by  orbifolding.
It is independent of the gauge group and the spacetime structure of
matter, as shown explicitly by our models with gauge groups $SU(3)
\times SU(2) \times U(1)$ and $SU(5)$, with matter in the bulk or on
the brane. The soft supersymmetry breaking interactions of squarks 
and sleptons do depend on whether matter is on the brane or in the 
bulk.  For the brane case, the squarks and sleptons are massless 
with $A = -\alpha/R$, Eq.~(\ref{eq:soft-scalar}), while in the 
bulk case the squarks and sleptons have degenerate mass-squareds 
$\alpha^2/R^2$ and $A = -3\alpha/R$, Eq.~(\ref{eq:soft-1}).

We have shown that this constrained form for the soft operators leads 
to successful electroweak symmetry breaking only in certain regions 
of parameter space.  For brane matter, $(1/R, \tan\beta)$ lie on a 
curve between $(2 \times 10^6~{\rm GeV}, 2)$ and 
$(6 \times 10^7~{\rm GeV}, 20)$; also, a large compactification 
scale $1/R \gsim 10^{14}~{\rm GeV}$ is allowed for low values of 
$\tan\beta \simeq 2$.  For bulk matter, the corresponding regions 
with successful electroweak symmetry breaking are 
$(700~{\rm GeV}, 2)$ to $(2~{\rm TeV}, 20)$, and, for larger 
compactification scales $(10^7~{\rm GeV}, 2)$ to 
$(10^{16}~{\rm GeV}, 4)$.  The unified theory prefers the case of 
bulk matter, since it gives electroweak symmetry breaking with 
less fine tuning.

Our supersymmetry breaking mechanism solves both the supersymmetric 
flavor and CP problems --- by construction the phases $\alpha$ and 
$\gamma$ are real and flavor conserving.  If matter is on the brane, 
the squark and slepton masses arise from renormalization group 
scaling with flavor blind gauge interactions.  With matter in the 
bulk, the orbifold symmetries necessarily lead to squark and 
slepton mass matrices proportional to the unit matrix.  
Non-trivial flavor mixing boundary conditions are inconsistent. 
However, flavor and CP violating scalar mass matrices could result 
in the case of bulk matter with large brane kinetic terms.

With an $SU(5)$ gauge group, the orbifold symmetry can be taken to be 
\begin{equation}
  U = e^{2\pi i \alpha T} \otimes P
\label{eq:orbifoldsym2}
\end{equation}
where $P$ is the parity $(-,-,-,+,+)$ acting on the ${\bf 5}$ of 
$SU(5)$. This single orbifold symmetry now breaks $SU(5)$ to the 
standard model gauge group, as well as breaking $G_H$ and supersymmetry.
We have explicitly constructed the unique 5d theory which accomplishes
this --- the only variations being the location of the matter multiplets. 

In this viewpoint the hierarchy problem is transformed to the question
of the origin of the small non-zero value of the orbifold mixing angle
$\alpha$.  The spacetime geometry is not fixed, but is ultimately
controlled by certain background fields, and the solution to the
hierarchy problem must be sought in the dynamics of these fields.

\section*{Acknowledgements}

Y.N. thanks the Miller Institute for Basic Research in Science 
for financial support.
This work was supported by the ESF under the RTN contract 
HPRN-CT-2000-00148, the Department of Energy under contract 
DE-AC03-76SF00098 and the National Science Foundation under 
contract PHY-95-14797.

\end{document}